\documentclass[aps,prd,onecolumn,groupedaddress,showpacs,nofootinbib,amssymb]{revtex4-2}
\usepackage[dvips]{graphicx}
\usepackage{amssymb}
\usepackage{amsmath}
\usepackage{graphicx,,color}
\usepackage{amsfonts}
\usepackage{bm}
\usepackage{cancel}
\usepackage{comment}
\usepackage{floatflt}
\newcommand\be{\begin{equation}}
\newcommand\ee{\end{equation}}

\allowdisplaybreaks[4]

\begin{document}

\tolerance=5000

\title{Chirality of Gravitational Waves in Chern-Simons $f(R)$ Gravity Cosmology}
\author{S.D. Odintsov,$^{1,2,3}$\,\thanks{odintsov@ice.cat}
V.K. Oikonomou,$^{4}$\,\thanks{v.k.oikonomou1979@gmail.com}}
\affiliation{$^{1)}$ ICREA, Passeig Luis Companys, 23, 08010 Barcelona, Spain\\
$^{2)}$ Institute of Space Sciences (ICE,CSIC) C. Can Magrans s/n,
08193 Barcelona, Spain\\
$^{3)}$ Institute of Space Sciences of Catalonia (IEEC),
Barcelona, Spain\\
$^{4)}$ Department of Physics, Aristotle University of
Thessaloniki, Thessaloniki 54124, Greece}
%$^{5)}$Laboratory for Theoretical Cosmology, International Center
%of Gravity and Cosmos, Tomsk State University of Control Systems
%and Radioelectronics  (TUSUR), 634050 Tomsk, Russia}

%\author{V.K.~Oikonomou,$^{1,2}$}
%\email{v.k.oikonomou1979@gmail.com,voikonomou@auth.gr}
%\affiliation{$^{1)}$ Department of Physics, Aristotle University
%of Thessaloniki, Thessaloniki 54124,
%Greece\\
%$^{2)}$ Laboratory for Theoretical Cosmology, Tomsk State
%University of Control Systems and Radioelectronics, 634050 Tomsk,
%Russia (TUSUR)\\
%}

 \tolerance=5000

\begin{abstract}
In this paper we shall consider an axionic Chern-Simons corrected
$f(R)$ gravity theoretical framework, and we shall study the
chirality of the generated primordial gravitational waves.
Particularly, we shall consider two main axion models, the
canonical misalignment axion model and the kinetic axion model,
both of which provide an interesting particle phenomenology, in
the presence of $R^2$ terms in the inflationary Lagrangian. The
axion does not affect significantly the background evolution
during the inflationary era, which is solely controlled by $R^2$
gravity. However, the due to the presence of the Chern-Simons
term, the tensor perturbations are directly affected, and our aim
is to reveal the extent of effects of the Chern-Simons term on the
gravitational waves modes, for both the axion models. We aim to
produce analytical descriptions of the primordial tensor modes
behavior, and thus we solve analytically the evolution equations
of the tensor modes, for a nearly de Sitter primordial evolution
controlled by the $R^2$ gravity. We focus the analytical study on
superhorizon and subhorizon modes. For the misalignment model, we
were able to produce analytic solutions for both the subhorizon
and superhorizon modes, in which case we found the behavior of the
circular polarization function. Our results indicate that the
produced tensor spectrum is strongly chiral. For the kinetic axion
model though, analytic solutions are obtained only for the
superhorizon modes. In order to have a grasp of the behavior of
the chirality of the tensor modes, we studied the chirality of the
superhorizon modes, however a more complete study is needed, which
is impossible to do analytically though.
\end{abstract}

%PACS numbers: 04.50.Kd, 95.36.+x, 98.80.-k, 98.80.Cq

\maketitle

\section{Introduction}

The next two decades are expected to be fascinating
scientifically, mainly because several interferometer experiments
are going to probe directly the existence or non-existence of
primordial stochastic gravitational waves
\cite{Baker:2019nia,Smith:2019wny,Seto:2001qf,Kawamura:2020pcg,Hild:2010id,Crowder:2005nr,Smith:2016jqs}.
These interferometers will probe directly frequencies that
correspond to stochastic tensor modes which re-entered the Hubble
horizon after inflation, during the early stages of the radiation
domination era. Thus these modes will reveal the physics of the
dark era, as is the reheating/radiation domination era is usually
dubbed. Indeed the physics of the dark era is unreachable by
terrestrial acceleration experiments, since the frequencies of the
aforementioned gravitational wave experiments correspond to
temperatures beyond and far beyond the temperature of the
electroweak phase transition. Apart from the aforementioned
interferometer experiments, there are also two experiments probing
intermediate frequencies, the Square Kilometer Array (SKA)
\cite{Bull:2018lat} which will soon start to give data, and the
pulsar timing arrays based NANOGrav collaboration
\cite{Arzoumanian:2020vkk,Pol:2020igl}. In the literature there
exist many theoretical descriptions of primordial gravitational
waves, see for example
\cite{Kamionkowski:2015yta,Denissenya:2018mqs,Turner:1993vb,Boyle:2005se,Kuroyanagi:2008ye,Smith:2005mm,
Liu:2015psa,Zhao:2013bba,Vagnozzi:2020gtf,Watanabe:2006qe,Kamionkowski:1993fg,Giare:2020vss,Zhao:2006mm,Zhao:2006eb,Cheng:2021nyo,Chongchitnan:2006pe,
Nakayama:2008wy,Capozziello:2017vdi,Capozziello:2008fn,Capozziello:2008rq,Cai:2021uup,Cai:2018dig,Odintsov:2021kup,Benetti:2021uea,Lin:2021vwc,Zhang:2021vak,Odintsov:2021urx,Breitbach:2018ddu}.
If a primordial stochastic tensor background is verified by the
interferometers, this will be a smoking gun for inflation
\cite{inflation1,inflation2,inflation3,inflation4}, which is the
most appealing and prominent scenario for the description of the
post-Planck evolution of our Universe. Traditionally, inflation is
described by scalar fields, however a viable alternative
description comes from modified gravity theories
\cite{reviews1,reviews2,reviews3,reviews4,reviews5,reviews6}. The
most important modified gravity theory is $f(R)$ gravity, which
allows in some cases a unified description of inflation with the
various subsequent evolution eras, like dark energy, see the
pioneer work \cite{Nojiri:2003ft} and also Refs.
\cite{Nojiri:2007cq,Cognola:2007zu,Nojiri:2006gh,Appleby:2007vb,Elizalde:2010ts,Odintsov:2020nwm,Oikonomou:2020qah,Oikonomou:2020oex}
for later developments along this research line.

In view of the exciting gravitational wave oriented next two
decades, in this article we aim to study the chirality of
primordial tensor modes in the context of an axionic Chern-Simons
corrected $f(R)$ gravity, the inflationary aspects of which were
studied in Ref. \cite{Odintsov:2019mlf}. The motivation to study
such axionic inflationary Lagrangians mainly comes from the fact
that the axion, or axion like particles, are quite appealing
candidates for particle dark matter, see
\cite{Marsh:2015xka,Marsh:2017yvc}, and also
\cite{Sikivie:2006ni,Raffelt:2006cw,Linde:1991km,maxim,Aoki:2017ehb}.
To our opinion, the axion is the last resort of particle dark
matter, but the predicted mass is too tiny to be measured at
present day. We shall consider two mainstream axion models, the
canonical misalignment model \cite{Marsh:2017yvc} and the recently
developed kinetic axion model
\cite{Co:2019jts,Co:2020dya,Barman:2021rdr}. Chern-Simons theories
are possible candidate theories toward describing the primordial
era of our Universe, see Refs.
\cite{Hwang:2005hb,Nishizawa:2018srh,Wagle:2018tyk,Yagi:2012vf,Yagi:2012ya,
Molina:2010fb,Izaurieta:2009hz,Sopuerta:2009iy,Konno:2009kg,Smith:2007jm,Matschull:1999he,
Haghani:2017yjk,Satoh:2007gn,Satoh:2008ck,Yoshida:2017cjl,Choi:1999zy,Satoh:2007gn,Satoh:2010ep,Odintsov:2019mlf}
and references therein, and see also \cite{Hang:2021oso} for
Chern-Simons topological terms. The Chern-Simons corrections in
the inflationary Lagrangian, have no direct effect on the
background evolution, however these do affect the tensor
perturbations, generating a chiral spectrum. For our study we
shall assume that the $f(R)$ gravity consists of the $R^2$ model
\cite{Starobinsky:1982ee}, and we shall study the evolution of the
primordial gravitational waves in the presence of a misalignment
or kinetic axion with Chern-Simons corrections. We aim to obtain
analytic solutions, so we shall solve analytically the evolution
equation for the tensor modes, focusing on superhorizon and
subhorizon modes. Using the resulting solutions, we study
quantitatively the effect of the Chern-Simons term by examining
the behavior of the circular polarization function. In both cases
we found that the tensor modes of $f(R)$ gravity with Chern-Simons
corrections are strongly chiral, as in the ordinary
Einstein-Hilbert case, however, for the kinetic axion we only
studied the superhorizon modes, due to the lack of analyticity.

This article is organized as follows: In section II, we present
the essential features of the Chern-Simons axionic $f(R)$ gravity
theoretical framework. We shall derive the field equations and
show explicitly how the Chern-Simons term affects the evolution of
the primordial tensor perturbations, while it does not affect at
all the background evolution and the scalar perturbations. We will
also present in brief the essential features of the misalignment
and kinetic axion theories, and we will show how the axion models
do not affect the inflationary era, which is controlled by the
$R^2$ gravity. In section III we thoroughly study in an analytic
way the evolution of the superhorizon and subhorizon modes for
both the two aforementioned axion models, and we will also
explicitly show in a quantitative way that the chirality feature
occurs even in the $f(R)$ gravity case. The conclusions of our
study is presented in the last section.

Before we start our analysis, we need to mention that the
geometric background which shall be assumed for the whole study is
a flat Friedmann-Robertson-Walker (FRW), with line element,
\begin{equation}
\label{metricfrw} ds^2 = - dt^2 + a(t)^2 \sum_{i=1,2,3}
\left(dx^i\right)^2\, ,
\end{equation}
where $a(t)$ is the scale factor. Also for the flat FRW metric,
the Hubble rate is $H=\frac{\dot{a}}{a}$ and the Ricci scalar is
$R=12H^2+6\dot{H}$.

\section{Essential Features of Chern-Simons Axion $f(R)$ Gravity, the $R^2$-Model and Two Axion Models}

We shall consider an $f(R)$ gravity in the presence of an axion
field with Chern-Simons term in vacuum, since we are interested in
primordial gravitational waves, which are generated during the
inflationary era. Thus we can safely ignore the radiation fluids,
and also any dark matter  perfect fluid components. In all modern
axionic models, the axion field plays the role of dark matter soon
after it starts to oscillate, when its mass  $m_a$ satisfies
$m_a\succeq H$, where $H$ is the Hubble rate. With regard to the
axion field, there exist in the literature two models which yield
viable phenomenology, the canonical misalignment axion model
\cite{Marsh:2015xka} and the kinetic axion model
\cite{Co:2019jts,Co:2020dya}. Both models may yield
phenomenologically appealing results, and for both models, when
$m_a\succeq H$, the axion field oscillations commence, beyond
which the axion field's energy density $\rho_a$ redshifts as dark
matter $\rho_a\sim a^{-3}$. The difference between the two axion
models is the time for which the axion oscillations commence, and
specifically for the case of the kinetic axion, the time instance
for which the axion oscillations start is significantly delayed
compared to the canonical misalignment model. In the kinetic axion
picture, the axion oscillations commence at some point during the
reheating era, at a lower temperature compared to the canonical
misalignment axion model. We shall discuss more these two axion
models later on in this section.

The whole $f(R)$ gravity axion model with Chern-Simons terms is a
sort of an $f(R,\phi)$ gravity in vacuum, the gravitational action
of which is,
\begin{equation}
\label{mainaction} \mathcal{S}=\int d^4x\sqrt{-g}\left[
\frac{1}{2\kappa^2}f(R)-\frac{1}{2}\partial^{\mu}\phi\partial_{\mu}\phi-V(\phi)+\frac{1}{8}\nu
(\phi)R\tilde{R} \right]\, ,
\end{equation}
with $R\tilde{R}=\epsilon^{abcd}R_{ab}^{ef}R_{cdef}$,
$\kappa^2=\frac{1}{8\pi G}$, where $G$ is Newton's gravitational
constant, and $\epsilon^{abcd}$ stands for the totally
antisymmetric Levi-Civita tensor. To be precise, the
Chern-Pontryagin density $R\tilde{R}$ is a direct analogue of the
term $ ^*F_{\mu \nu}F^{\mu \nu}$ in principal fibre bundles, which
is constructed by the curvature $F_{\mu \nu}$ which corresponds to
the connection $A_{\mu}$ of the principal bundle. In the
literature the term $\nu (\phi)\tilde{R}R$ is called Chern-Simons
term, but this terminology is abusively used for a simple reason,
the Chern-Pontryagin density $\nu (\phi)\tilde{R}R$ is directly
connected with an actual 3d Chern-Simons term cohomologicaly $\nu
(\phi)\tilde{R}R=d(Chern-Simons)$, via the exterior derivative.
This is the justification of the Chern-Simons terminology, which
we shall also adopt in this paper, complying with the literature.
In the context of the metric formalism, upon varying the
gravitational action (\ref{mainaction}) with respect to the metric
tensor and with respect to the scalar field respectively, and for
the FRW metric (\ref{metricfrw}), the following equations of
motion are obtained,
\begin{align}\label{eqnsofmkotion}
& 3 H^2F=\kappa^2\frac{1}{2}\dot{\phi}^2+\frac{RF-f+2
V\kappa^2}{2}-3H\dot{F}\, ,\\ \notag &
-3FH^2+2\dot{H}F=\kappa^2\frac{1}{2}\dot{\phi}^2-\frac{RF-f+2
V}{2}+\ddot{F}+2H\dot{F}\, ,
\end{align}
\begin{equation}\label{scalarfieldeqn}
\ddot{\phi}+3H\dot{\phi}+V'(\phi)=0\, ,
\end{equation}
with $V'(\phi)=\frac{\partial V}{\partial \phi}$ and also
$F=\frac{\partial f}{\partial R}$.
\begin{figure}
\centering
\includegraphics[width=18pc]{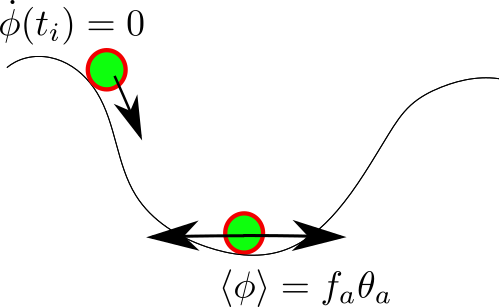}
\caption{The canonical misalignment axion physics. The axion
starts uphill from a small deviation from its vacuum expectation
value, with zero velocity, and it finally ends up in the minimum
where it starts oscillating when $H\sim m_a$.}\label{plot1}
\end{figure}
Remarkably, the field equations (\ref{eqnsofmkotion}) and
(\ref{scalarfieldeqn}), remain totally unaffected by the
Chern-Simons term, as it is also pointed out in the literature
\cite{Hwang:2005hb}, see also \cite{Choi:1999zy}. Basically, the
whole background evolution in the presence of the Chern-Simons
term remains entirely unaffected by the Chern-Simons term, and
only the tensor perturbations and the corresponding slow-roll
indices are affected by the Chern-Simons term \cite{Hwang:2005hb}
(see also \cite{Odintsov:2019mlf}), as we also see in the next
section. The reason for this is the fact it is impossible to have
$\epsilon^{abcd}$ and scalar derivatives only in any of the
following components of the energy momentum tensor, $T_{00}$,
$T_{0 \alpha}$, $T_{\alpha \beta}$ \cite{Choi:1999zy}.

At this point, let us specify the inflationary $f(R)$ gravity
model which we shall assume that controls the inflationary
evolution. Specifically, we shall assume that the $f(R)$ gravity
model which governs the evolution is the $R^2$ model
\cite{Starobinsky:1982ee},
\begin{equation}\label{starobinskyappendix}
f(R)=R+\frac{1}{36H_i}R^2\, ,
\end{equation}
where $H_i$ has mass dimensions $[m]^2$. As it was shown in Ref.
\cite{Odintsov:2019mlf}, the inflationary evolution is governed
mainly by the $R^2$ gravity, because the axion is frozen (it has
small deviations from the vacuum expectation value, a time average
is assumed) in its vacuum expectation value during inflation, and
due to its small mass, for example, the potential term is of the
order,
\begin{equation}\label{termlastleading}
\frac{\kappa^2}{2
(12H^2)}m_a^2f_a^2\theta_a^2=\mathcal{O}(10^{-39})eV\, ,
\end{equation}
with $H\sim H_I=\mathcal{O}(10^{13})$GeV, while the $R^2$-related
terms are of the order $\mathcal{O}(10^{38})$eV. Thus, by
substituting the Ricci scalar $R=12H^2+6\dot{H}$, its derivative
$\dot{R}=24H\dot{H}+6\ddot{H}$ and $F=1+\frac{R}{18H_i}$ in the
Friedmann equation (\ref{eqnsofmkotion}), we obtain,
\begin{equation}\label{patsunappendix}
\ddot{H}-\frac{\dot{H}^2}{2H}+3H_iH=-3H\dot{H}\, .
\end{equation}
Disregarding the first two terms during the slow-roll era, the
differential equation (\ref{patsunappendix}) becomes,
\begin{equation}\label{patsunappendix1}
3H_iH=-3H\dot{H}\, ,
\end{equation}
which yields the solution,
\begin{equation}\label{quasidesitter}
H(t)=H_0-H_i t\, ,
\end{equation}
which is a quasi-de Sitter evolution. The spectral index and the
tensor-to-scalar ratio for the combined axion Chern-Simons $R^2$
model are \cite{Odintsov:2019mlf},
\begin{equation}\label{spectralstarobinsky}
n_s=1-\frac{2}{N},\,\,\,r\simeq
\frac{r_s^v}{2}\left(\frac{1}{|1-\frac{\kappa^2x}{F}|}+\frac{1}{|1+\frac{\kappa^2x}{F}|}\right)\,
,
\end{equation}
with $r_s^v= 48\epsilon_1^2$ being the tensor-to-scalar ratio of
vacuum $f(R)$ gravity, and $\epsilon_1$ is the first slow-roll
index, which for the $R^2$ model is $\epsilon_1\sim 1/(2N)$. Also
the parameter $x$ is defined as $x=\frac{2\dot{\nu}k}{a}$. As it
was shown in \cite{Odintsov:2019mlf}, the term $\sim
\frac{\kappa^2x}{F}$ can potentially reduce the tensor-to-scalar
ratio, for example taking $\frac{\kappa^2x}{F}=\mathcal{O}(3\times
10^2)$, the tensor-to-scalar ratio becomes of the order $r\sim
\mathcal{O}(10^{-5})$. Now let us discuss in brief the two axion
models which we shall consider in this paper. The first is the
canonical misalignment model, which we now present in brief. In
the canonical misalignment model (see \cite{Marsh:2015xka} for a
review), the pre-inflationary Peccei-Quinn $U(1)$ symmetry which
was unbroken pre-inflationary, is broken during inflation, and the
axion field acquires a non-zero vacuum expectation value $\langle
\phi \rangle =\theta_a f_a$, where $f_a$ denotes the axion decay
constant, and $\theta_a$ denotes the initial misalignment angle.
The axion vacuum expectation value is quite large during the
inflationary period, of the order of the axion decay constant
$\sim \mathcal{O}(10^{10})$GeV. We need to stress that although
that the axion has a vacuum expectation value during the
inflationary era, this does not mean that the axion is constant
during inflation, meaning that the Chern-Simons term during
inflation should be viewed as a time average $\langle \bar{\nu
(\phi)}\tilde{R}R \rangle$, and its effect is non trivial
considering its time averaged values. The axion potential is,
\begin{equation}\label{axionpotentnialfull}
V(\phi )=m_a^2f_a^2\left(1-\cos (\frac{\phi}{f_a}) \right)\, .
\end{equation}
Now let us discuss how the canonical misalignment model functions,
which schematically appears in Fig. \ref{plot1}. The axion during
inflation has small displacements from its vacuum expectation
value and starts to roll down the hill with initial speed
$\frac{\dot{\phi}}{m_a}\ll 1$, so quite small or nearly zero
initial speed. For small displacements from the vacuum expectation
value, the potential is approximately equal to,
\begin{equation}\label{axionpotential}
V(\phi )\simeq \frac{1}{2}m_a^2\phi^2\, ,
\end{equation}
when $\phi\ll f_a$ or equivalently $\phi\ll \langle \phi \rangle
$. Hence, the axion rolls down from the hill when $H\gg m_a$, as
it is displayed in Fig. \ref{plot1}, until when $H\sim m_a$ at
which point it starts to oscillate and its energy density
redshifts as dark matter. Hence the misalignment model is based on
the fact that the initial speed of the axion at small
displacements from its vacuum expectation value on the hill of the
potential, the velocity and the acceleration are quite small.
Hence for the canonical misalignment axion model we have
$\frac{\dot{\phi}}{m_a}\ll 1$ and $\frac{\ddot{\phi}}{m_a^2}\ll
1$.
\begin{figure}
\centering
\includegraphics[width=18pc]{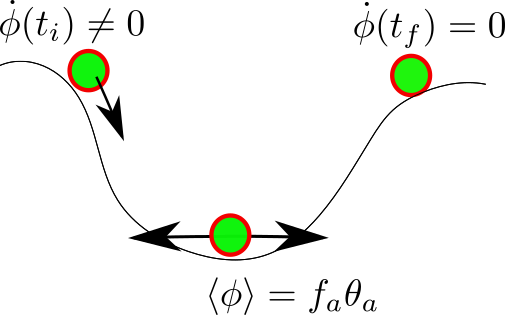}
\caption{The kinetic axion physics. The axion starts uphill from a
small deviation from its vacuum expectation value, with non-zero
velocity, and it finally ends up uphill again, from where it goes
again downhill until it ends up in the minimum, where it starts
oscillating when $H\sim m_a$. In this case, the oscillation period
is postponed compared to the canonical misalignment model, and it
occurs at a lower temperature.}\label{plot2}
\end{figure}
Now the kinetic axion model \cite{Co:2019jts,Co:2020dya} is based
on the fact that initial, when the axion was on the hill of the
potential, for small displacement from its vacuum expectation
value, the speed was not zero, as is seen in Fig. \ref{plot2}.
Thus as the axion rolls down the hill, it does not start to
oscillate as it reaches the minimum of the potential, but goes
uphill, and when it stops it rolls down to oscillate around the
vacuum expectation value, where it starts to redshift as dark
matter. For the kinetic axion case, initially the speed is quite
large, and specifically the axion kinetic energy dominates over
the potential energy \cite{Co:2019jts,Co:2020dya} $\dot{\phi}^2\gg
m_a^2\phi^2$, thus essentially equation of state parameter $w$ for
the axion initially, prior to the axion oscillations, and during
the whole inflationary era, is approximately $w\sim 1$, thus the
axion velocity redshifts as $\dot{\phi}\sim a^{-3}$ so this is a
stiff equation of state. As an effect of this initial kinetic
domination, the axion does not start its oscillations around its
vacuum expectation value, but continues uphill. For the kinetic
axion case, since the kinetic energy term dominates over the
potential, and in conjunction with the fact that $\dot{\phi}^2\sim
a^{-6}$, this simply means that the background evolution is solely
governed by $f(R)$ gravity. Hence, the $R^2$ gravity controls the
background evolution in this case too, as in the canonical
misalignment axion model, thus the evolution is the quasi-de
Sitter one of Eq. (\ref{quasidesitter}).

In the next section we shall consider the evolution of primordial
gravitational waves for the axion $R^2$ model, for both the axion
models we discussed in this section and we shall explore the
chirality of the produced gravitational waves in a semi-analytical
way.

\section{Primordial Gravitational Waves in Chern-Simons Axion $f(R)$ Gravity}

In this section we shall consider the evolution of primordial
gravitational waves in the context of axion $f(R)$ gravity with
Chern-Simons corrections. We shall concentrate on the $R^2$
gravity case, in which case as we showed earlier, for both the
canonical misalignment and the kinetic axion, the background
evolution is governed solely by the $R^2$ gravity.
\begin{figure}
\centering
\includegraphics[width=22pc]{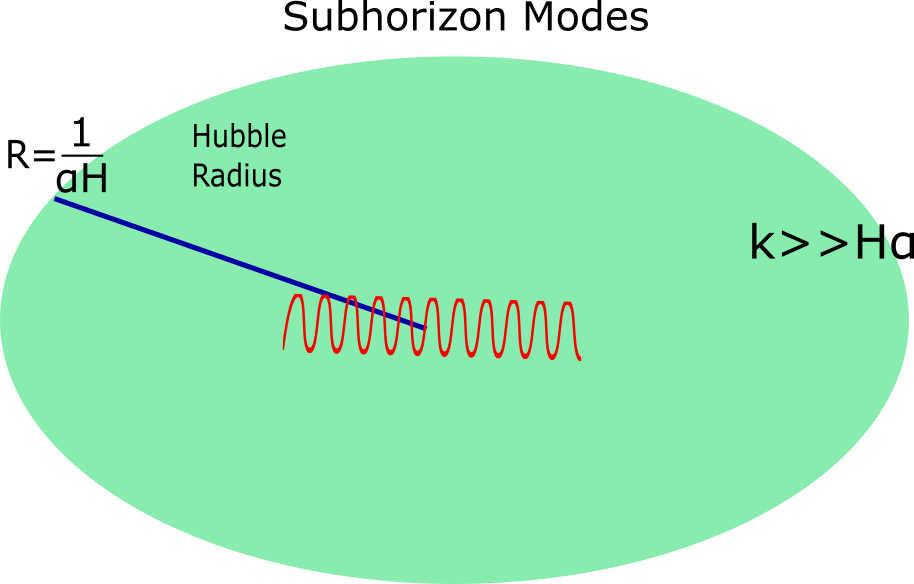}
\caption{Subhorizon inflationary modes. During inflation these are
at subhorizon scales, and the ones with the smallest wavenumber
will be the first that will exit the horizon after the
inflationary era ends and during the early stages of the reheating
era. These modes will be probed in the next years by  space laser
interferometers, like LISA.}\label{plot3}
\end{figure}
Let us now consider the evolution of tensor perturbations, and to
start of, we shall consider the following tensor perturbations of
the flat FRW metric,
\begin{equation}\label{tensorpert}
\mathrm{d}s^2=-\mathrm{d}t^2+a(t)^2\left(\delta_{ij}+h_{ij}
\right)\mathrm{d}x^i\mathrm{d}x^j\, ,
\end{equation}
and by performing the Fourier transform of the tensor perturbation
$h_{i j}$,
\begin{equation}\label{tensorperturbationfourier}
h_{i j}(\vec{x},t)=\sqrt{V}\int
\frac{\mathrm{d}^3k}{(2\pi)^3}\sum_{\ell}\epsilon_{i
j}^{\ell}h_{\ell k}e^{i\vec{k}\vec{x}}\, ,
\end{equation}
where ``$\ell$'' denotes the polarization of the tensor
perturbation, and $V$ is the volume element. The Fourier
transformation of the tensor perturbation $h_{i j}$ satisfies the
following differential equation \cite{Hwang:2005hb},
\begin{equation}\label{mainevolutiondiffeqnfrgravity}
\ddot{h}_{\ell}(k)+\left(3+\alpha_M
\right)H\dot{h}_{\ell}(k)+\frac{k^2}{a^2}h_{\ell}(k)=0\, ,
\end{equation}
where the parameter $\alpha_M$ is defined to be,
\begin{equation}\label{amfrgravity}
a_M=\frac{\dot{Q}_t}{Q_tH}\, ,
\end{equation}
and the function $Q_t$ for the Chern-Simons axion $f(R)$ gravity
is equal to \cite{Hwang:2005hb},
\begin{equation}\label{qttermchernsimons}
Q_t=\frac{1}{\kappa^2}\frac{\mathrm{d} f}{\mathrm{d}
R}+\frac{2\lambda_{\ell}\dot{\nu}k}{a}\, ,
\end{equation}
and the parameter $\lambda_{\ell}$ indicates the polarization of
the gravitational waves and it takes the following values,
$\lambda_{R}=1$ for the right handed gravitational waves, and
$\lambda_{L}=-1$ for the left handed gravitational waves. Also,
$k$ denotes the wavenumber each tensor mode. Easily we can
evaluate the exact form of the parameter $a_M$ for the
Chern-Simons-corrected axion $f(R)$ gravity has the following
form,
\begin{equation}\label{amfrphichernsimons}
a_M=\frac{\frac{1}{\kappa^2}\frac{\mathrm{d}^2f}{\mathrm{d}
R^2}\dot{R}+\frac{2\lambda_{\ell}\ddot{\nu}k}{a}-\frac{2\lambda_{\ell}\dot{\nu}k\,H}{a}}{\left(\frac{1}{\kappa^2}\frac{\mathrm{d}
f}{\mathrm{d} R}+\frac{2\lambda_{\ell}\dot{\nu}k}{a}\right)H}\, .
\end{equation}
Basically, the parameter $a_M$ quantifies the direct effect of the
modified gravity under study, since in the absence of this term,
the differential equation (\ref{mainevolutiondiffeqnfrgravity}) is
identical to the general relativistic differential equation
governing the primordial tensor perturbations.

\begin{figure}
\centering
\includegraphics[width=22pc]{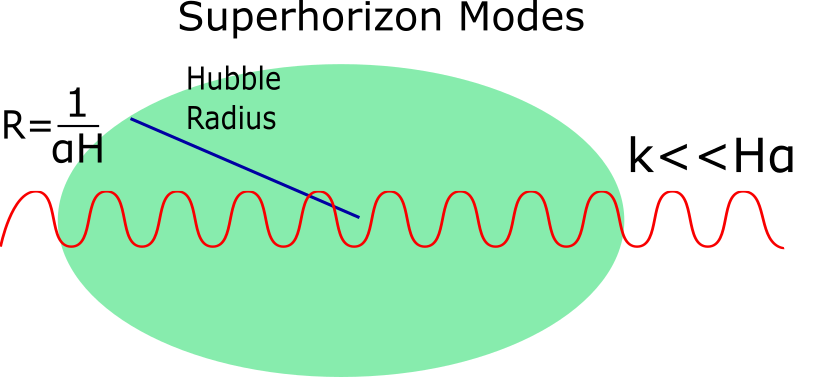}
\caption{Superhorizon modes, for which $\lambda\gg \frac{1}{a H}$,
or equivalently $k\ll a H$. The CMB modes were superhorizon very
early during the inflationary era. Our interest is in CMB
superhorizon modes, so for $\lambda < 10$Mpc, or equivalently with
$k<0.1$Mpc$^{-1}$.}\label{plot4}
\end{figure}

For inflationary gravitational waves, there are two cases of
interest, regarding the magnitude of the mode wavelength compared
to the Hubble horizon, the subhorizon and superhorizon modes. Both
modes are equally important from an experimental perspective. The
subhorizon modes during inflation and especially the ones with the
smallest wavenumber, will be the first that will exit the horizon
after inflation and during the early stages of the reheating era.
These subhorizon modes (see Fig. \ref{plot3}), for which their
wavelength is significantly smaller than the Hubble radius, that
is, $\lambda\ll \frac{1}{a H}$, or equivalently $k\gg a H$ will
directly be probed in about fifteen years from now, in LISA,
DECIGO BBO and other large frequency gravitational waves
experiments. Thus the study of these modes even analytically is of
some importance. In this paper we shall mainly be interested in
studying the chirality of these modes, in order to see
quantitatively and at first hand their evolution. With regards to
the superhorizon modes, for these their wavelength is quite larger
than the Hubble radius, that is, $\lambda\gg \frac{1}{a H}$, or
equivalently $k\ll a H$ (see Fig. \ref{plot4}).
These modes will not be probed by the space interferometers, but
are relevant because the CMB modes are basically modes that became
superhorizon very early during the inflationary era, and were
superhorizon until $z\sim 1100$, where they reentered the Hubble
horizon. We shall be interested in CMB superhorizon modes, so
basically for modes with wavelength $\lambda < 10$Mpc, or
equivalently with $k<0.1$Mpc$^{-1}$. With regard to the subhorizon
modes, we shall be interested for wavelengths larger than $\lambda
>10$Mpc or equivalently for $0.1<k<10^{16}$Mpc$^{-1}$. The modes
with $0.1<k<10^{16}$Mpc$^{-1}$ correspond to post-CMB and relevant
to future primordial gravitational wave experiments, especially
the modes with $k>10^{10}$Mpc$^{-1}$. The modes with
$10^{4}<k<10^{8}$Mpc$^{-1}$ will be probed by the SKA or NANOGrav
collaborations. For the above reasons, in this work we shall
analytically check in a quantitative way the chirality of both
subhorizon and superhorizon modes for both the axion $R^2$ gravity
models.

Before we specify our analysis using the two axion $R^2$ models we
discussed previously, let us consider the general behavior of the
superhorizon modes in a model independent way. For the
superhorizon modes, the differential equation
(\ref{mainevolutiondiffeqnfrgravity}) can be greatly simplified,
since for these modes, their wavelengths are significantly larger
compared the Hubble horizon, that is $k\ll H a$, so basically the
third term in the differential equation
(\ref{mainevolutiondiffeqnfrgravity}) can be safely omitted, thus
it reads,
\begin{equation}\label{mainevolutiondiffeqnfrgravityup}
\ddot{h}_{\ell}(k)+\left(3+\alpha_M \right)H\dot{h}_{\ell}(k)=0\,
.
\end{equation}
We can find a general solution of the above differential equation,
which reads,
\begin{equation}\label{generalsolutiondiffeqn}
h_{\ell}(k)=C_{\ell}(k)+D_{\ell}(k)\int_1^t \exp
\left(\int_1^{\eta} (-a_M(\tau)-3 H(\tau)) \,
\mathrm{d}\tau\right) \, \mathrm{d}\eta \, ,
\end{equation}
thus the solution describes clearly a time-independent frozen term
$C_{\ell}(k)$ and an exponentially decaying mode, which is the
second term in Eq. (\ref{generalsolutiondiffeqn}). We shall
explicitly verify this behavior for both the canonical
misalignment and kinetic axion $R^2$ models later on in this
section.

Now our analysis will be to reveal in an analytic way the amount
of chirality of the primordial gravitational waves relevant for
both the CMB and future gravitational waves experiments, thus both
for superhorizon and subhorizon modes, where analytical results
can be obtained of course. Our strategy will be the following,
since we are interested for the inflationary gravitational waves
for the axion $R^2$ gravity models, we shall solve analytically
the evolution differential equation
(\ref{mainevolutiondiffeqnfrgravity}), for both the left handed
and right handed polarizations, considering both the superhorizon
and subhorizon modes. Suppose the solutions are $h_{L}(k)$ and
$h_{R}(k)$, and these solutions contain two integration constants
each. When analytic results can be obtained, in order to determine
the integration constants, we shall assume that both solutions
asymptotically in the past satisfy the Bunch-Davies initial
condition, and we shall find the rest of the constants in an easy
way by taking the asymptotic expansions of the solutions in two
regimes, the subhorizon and superhorizon regimes, so the solutions
would be $h_{L}(k)_{k\ll a H}$, $h_{L}(k)_{k\gg a H}$,
$h_{R}(k)_{k\ll a H}$ and $h_{R}(k)_{k\ll a H}$. Then the
integrations constants can be obtained by matching the solutions
for each polarization at an intermediate transition time
$t=t_{trans}$, as follows,
\begin{align}\label{intermediateinitialconditions}
& h_{L}(k)_{k\ll a H}\Big{|}_{t=t_{trans}}=h_{L}(k)_{k\gg a
H}\Big{|}_{t=t_{trans}},
\\ \notag &
\dot{h}_{L}(k)_{k\ll a
H}\Big{|}_{t=t_{trans}}=\dot{h}_{L}(k)_{k\gg a
H}\Big{|}_{t=t_{trans}},\\ \notag & h_{R}(k)_{k\ll a
H}\Big{|}_{t=t_{trans}}=h_{R}(k)_{k\gg a
H}\Big{|}_{t=t_{trans}},
\\ \notag &
\dot{h}_{R}(k)_{k\ll a
H}\Big{|}_{t=t_{trans}}=\dot{h}_{R}(k)_{k\gg a
H}\Big{|}_{t=t_{trans}}\, .
\end{align}
With the above initial conditions
(\ref{intermediateinitialconditions}), the integration constants
can be obtained, when both the superhorizon and subhorizon modes
can be obtained. As we shall see, for the canonical misalignment
axion $R^2$ model, this is possible, however for the kinetic axion
$R^2$ model, analytic solutions can be obtained only for the
superhorizon modes. Thus for the sake of the argument, we shall
use arbitrary integration constants only for this case, just to
see the behavior of the solutions as functions of the wavenumber
of the modes, and also in order to reveal quantitatively the
chirality between the left and right handed modes. For the
analysis of the chirality between the left and right handed modes,
which quantifies the difference in the propagation between left
handed and right handed modes, we shall share the circular
polarization function $\Pi(k)$ from electromagnetic studies,
defined as $\Pi(k)=\frac{Q}{I}$, where $Q$ an $I$ are the stokes
parameters for electromagnetic waves, defined as
$I=|E_{x}|^2+|E_{y}|^2$, $Q=|E_{x}|^2-|E_{y}|^2$
\cite{Alexander:2018iwy}. Thus for the gravitational waves, the
circular polarization function $\Pi(k)$ is defined as follows
\cite{Satoh:2008ck},
\begin{equation}\label{circularpolarizationformulasub}
\Pi_{k\gg a H}(k)=\frac{|h_{L}^{k\gg a H}(k)|^2-|h_{R}^{k\gg a
H}(k)|^2}{|h_{L}^{k\gg a H}(k)|^2+|h_{R}^{k\gg a H}(k)|^2}\, ,
\end{equation}
for the case of subhorizon modes, while for superhorizon modes, we
have,
\begin{equation}\label{circularpolarizationformulasup}
\Pi_{k\ll a H}(k)=\frac{|h_{L}^{k\ll a H}(k)|^2-|h_{R}^{k\ll a
H}(k)|^2}{|h_{L}^{k\ll a H}(k)|^2+|h_{R}^{k\ll a H}(k)|^2}\, .
\end{equation}
Our aim in the rest of this section is to reveal the behavior of
the circular polarization functions $\Pi_{k\gg a H}(k)$ and
$\Pi_{k\ll a H}(k)$ for both the kinetic and canonical
misalignment axion $R^2$ gravity, when this is possible.

Let us start with the canonical misalignment axion $R^2$ gravity.
In this scenario, as we mentioned earlier, the kinetic energy and
the acceleration of the axion field, namely $\dot{\phi}$ and
$\ddot{\phi}$, are insignificant, that is
$\frac{\dot{\phi}}{m_a}\ll 1$, $\frac{\ddot{\phi}}{m_a^2}\ll 1$,
and also note that during the inflationary era $m_a\ll H$. Thus we
can fix these to have significantly small values compared to the
axion mass, which we shall assume that it is of the order $m_a\sim
\mathcal{O}(10^{-12})$eV for the canonical misalignment case. In
view of the above considerations, by also taking into account that
$\ddot{\phi}\ll H \dot{\phi}$ for the misalignment axion, we shall
take $\dot{\phi}\sim \mathcal{O}(10^{-5}m_a$ and $\ddot{\phi}\sim
\mathcal{O}(10^{-5}m_a^2)$. Also for the rest of this paper, we
shall assume that the Chern-Simons coupling function $\nu (\phi)$
has the form,
\begin{equation}\label{nuphi}
\nu(\phi)=\gamma e^{\beta \phi \kappa}\, ,
\end{equation}
where recall, $\kappa=\frac{1}{M_p}$, where $M_p$ is the reduced
Planck mass. The values of the dimensionless free parameters
$\gamma$ and $\beta$ are determined by using the rule to have a
viable inflationary phenomenology. We shall return to this issue
later on, but now let us simplify the evolution equation
(\ref{mainevolutiondiffeqnfrgravity}) as much as possible in order
to have a concrete quantitative idea on the behavior of the
circular polarization functions $\Pi_{k\gg a H}(k)$ and $\Pi_{k\ll
a H}(k)$ for both the subhorizon and superhorizon modes
respectively. In both cases, since the modes are inflationary
modes, the derivative of the $f(R)$ gravity can be simplified as
$F(R)\sim \alpha R$, where $\alpha=\frac{1}{18H_i}$, and also the
Ricci scalar is approximately $R\sim 12 H_0^2$. In addition, since
the evolution is quasi-de Sitter, we assume that it is exactly a
de Sitter evolution. Using these simplifications, and due to the
fact that $\frac{\ddot{\nu}}{\kappa}\ll \dot{\nu}$, the simplified
evolution equation (\ref{mainevolutiondiffeqnfrgravity}) for the
left-handed polarization subhorizon modes reads,
\begin{equation}\label{lefthandedevo}
\ddot{u}_L (t)+\dot{u}_L(t) \left(\frac{k \dot{a}(t)
\dot{\nu}(t)}{12 \alpha H_0^2 a(t)^2}+3 H_0\right)+\frac{k^2
u_L(t)}{a(t)^2}=0\, ,
\end{equation}
while the right-handed polarization subhorizon modes satisfy,
\begin{equation}\label{righthandedevo}
\ddot{u}_R (t)+\dot{u}_R(t) \left(-\frac{k \dot{a}(t)
\dot{\nu}(t)}{12 \alpha H_0^2 a(t)^2}+3 H_0\right)+\frac{k^2
u_R(t)}{a(t)^2}=0\, .
\end{equation}
The two equations can be solved analytically by simply taking into
account the slowly varying form of $\dot{\nu}$ we discussed
earlier. By taking a constant value for $\dot{\phi}=\dot{\phi}\nu
'(\phi)=\delta $, which we shall consider its values later on, by
solving the differential equation (\ref{lefthandedevo}) the
left-handed evolution function $u_L(t)$ has the following form,
\begin{align}\label{lefthandedsolution}
& u_L(t)=\mathcal{C}_1 U\left(-\frac{-\delta -2 \sqrt{\delta
^2-576 H_0^2 \alpha ^2}}{\sqrt{\delta ^2-576 H_0^2 \alpha
^2}},4,\frac{e^{-H_0 t} k \sqrt{\delta ^2-576
H_0^2 \alpha ^2}}{12 H_0^2 \alpha }\right) \times \\
\notag &\exp \left(\frac{72 \alpha  H_0^2 \log \left(e^{-H_0
t}\right)-k e^{-H_0 t} \left(\sqrt{\delta ^2-576 \alpha ^2
H_0^2}-\delta \right)}{24 \alpha  H_0^2}\right) \\
\notag &+\mathcal{C}_2 L_{n}^3\left(\frac{k e^{-H_0 t}
\sqrt{\delta ^2-576 \alpha ^2 H_0^2}}{12 \alpha H_0^2}\right) \exp
\left(\frac{72 \alpha  H_0^2 \log \left(e^{-H_0 t}\right)-k
e^{-H_0 t} \left(\sqrt{\delta ^2-576 \alpha ^2 H_0^2}-\delta
\right)}{24 \alpha  H_0^2}\right)\, ,
\end{align}
where $U(a,b,z)$ is the confluent hypergeometric function and
$L_n^a(x)$ is the generalized Laguerre polynomial, and
$n=\frac{-\delta -2 \sqrt{\delta ^2-576 \alpha ^2
H_0^2}}{\sqrt{\delta ^2-576 \alpha ^2 H_0^2}}$. Also
$\mathcal{C}_1$ and $\mathcal{C}_2$ are integration constants that
will be determined by the initial conditions
(\ref{intermediateinitialconditions}), after we will obtain the
superhorizon solutions. Accordingly, by solving the right-handed
modes equation (\ref{righthandedevo}), we obtain the right-handed
evolution function $u_R(t)$ which has the following form,
\begin{align}\label{righthandedsolution}
& u_R(t)=\mathcal{C}_3 U\left(-\frac{\delta -2 \sqrt{\delta ^2-576
H_0^2 \alpha ^2}}{\sqrt{\delta ^2-576 H_0^2 \alpha
^2}},4,\frac{e^{-H_0 t} k \sqrt{\delta ^2-576 H_0^2
\alpha ^2}}{12 H_0^2 \alpha }\right) \\
\notag & \exp \left(\frac{72 \alpha  H_0^2 \log \left(e^{-H_0
t}\right)-k e^{-H_0 t} \left(\delta +\sqrt{\delta ^2-576 \alpha ^2
H_0^2}\right)}{24 \alpha
H_0^2}\right)\\
\notag & +\mathcal{C}_4 L_{b}^3\left(\frac{k e^{-H_0 t}
\sqrt{\delta ^2-576 \alpha ^2 H_0^2}}{12 \alpha H_0^2}\right) \exp
\left(\frac{72 \alpha  H_0^2 \log \left(e^{-H_0 t}\right)-k
e^{-H_0 t} \left(\delta +\sqrt{\delta ^2-576 \alpha ^2
H_0^2}\right)}{24 \alpha H_0^2}\right)\, ,
\end{align}
with $b=\frac{\delta -2 \sqrt{\delta ^2-576 \alpha ^2
H_0^2}}{\sqrt{\delta ^2-576 \alpha ^2 H_0^2}}$. Also
$\mathcal{C}_3$ and $\mathcal{C}_4$ are integration constants that
in this case too will be determined by the initial conditions
(\ref{intermediateinitialconditions}). Now let us turn our focus
on superhorizon modes.
\begin{figure}
\centering
\includegraphics[width=20pc]{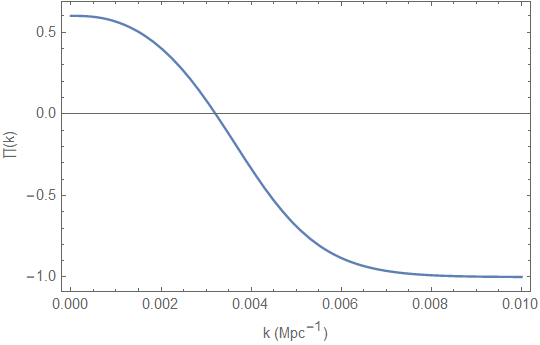}
\includegraphics[width=20pc]{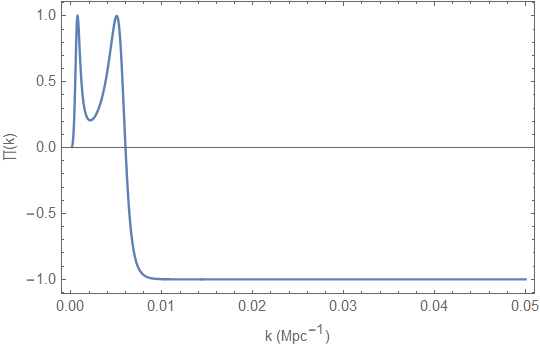}
\caption{The circular polarization function $\Pi_{k\gg a
H}(k)=\frac{|u_{L}^{k\gg a H}(k)|^2-|u_{R}^{k\gg a
H}(k)|^2}{|u_{L}^{k\gg a H}(k)|^2+|u_{R}^{k\gg a H}(k)|^2}$ for
the subhorizon modes (left plot), for the canonical misalignment
axion model, as a function of the wavenumber of the modes. The
right plot represents the circular polarization function
$\Pi_{k\ll a H}(k)=\frac{|u_{L}^{k\ll a H}(k)|^2-|u_{R}^{k\ll a
H}(k)|^2}{|u_{L}^{k\ll a H}(k)|^2+|u_{R}^{k\ll a H}(k)|^2}$ for
the superhorizon modes, as function of the wavenumber of the
modes.}\label{plot5}
\end{figure}
Regarding these modes, the evolution differential equation can
easily be solved analytically, just in the previous case. These
modes are relevant for CMB observations, and we now will measure
the polarization of the superhorizon modes for small values of
$k$, in the range $10^{-4}\leq k \leq 10$Mpc$^{-1}$. From the
beginning though, we know that superhorizon modes do freeze once
they become superhorizon, so we expect a decay of the functions
$u_L(t)$ and $u_R(t)$ after the horizon crossing, as functions of
time, and in fact an exponential decay. We shall present the
detailed study of the superhorizon modes here, so regarding the
left-handed polarization superhorizon modes, the differential
equation governing their evolution is basically the same as in Eq.
(\ref{lefthandedevo}) by simply omitting the last term, so in this
case the superhorizon function $u_L(t)$ which we shall denote as
$u_L^s(t)$ has the following form,
\begin{align}\label{lefthandedsolutionsuper}
& u_L^s(t)=\mathcal{C}_5+\mathcal{C}_6 e^{\frac{\delta  k e^{-H_0
t}}{12 \alpha H_0^2}} \left(-\frac{3456 \alpha ^3 H_0^5}{\delta ^3
k^3}+\frac{288 \alpha ^2 H_0^3 e^{-H_0 t}}{\delta ^2 k^2}-\frac{12
\alpha  H_0 e^{-2 H_0 t}}{\delta k}\right)\, ,
\end{align}
where $\mathcal{C}_5$ and $\mathcal{C}_6$ are integration
constants that in this case too will be determined by the initial
conditions (\ref{intermediateinitialconditions}). As expected, the
superhorizon left-handed modes essentially freeze after the
horizon crossing and they also exponentially decay as functions of
the cosmic time.  Also for the righthanded polarization modes, the
corresponding solution $u_R^s(t)$ reads,
\begin{align}\label{righthandedsolutionsuper}
& u_R^s(t)=\mathcal{C}_7+\mathcal{C}_8 e^{-\frac{\delta  k e^{-H_0
t}}{12 \alpha H_0^2}} \left(\frac{3456 \alpha ^3 H_0^5}{\delta ^3
k^3}+\frac{288 \alpha ^2 H_0^3 e^{-H_0 t}}{\delta ^2 k^2}+\frac{12
\alpha  H_0 e^{-2 H_0 t}}{\delta k}\right)\, ,
\end{align}
where in this case too $\mathcal{C}_5$ and $\mathcal{C}_6$ are
integration constants that in this case too will be determined by
the initial conditions (\ref{intermediateinitialconditions}).
Also, the right-handed solution decays exponentially in time too.
Now by using the initial conditions
(\ref{intermediateinitialconditions}), and the Bunch-Davies
initial condition for each mode we can find the constants of
integration, which we omit for brevity. In a previous section we
saw that for the model under consideration, a viable phenomenology
is achieved for $\frac{\kappa^2x}{F}=\mathcal{O}(3\times 10^2)$.
This can be easily arranged for various values of the free
parameters, for both the superhorizon and subhorizon modes, at the
first horizon crossing though. By taking these into account, and
also the values of the integration constants, and finally by
considering inflationary times of the order $t\sim 10^{-30}$sec,
in Fig. \ref{plot5} we present the behavior of the circular
polarization function $\Pi_{k\gg a H}(k)$ for modes that were
subhorizon during the first stages of inflation. These subhorizon
modes are directly relevant to the future gravitational waves
experiments, so the wavenumber should be from $k=10$Mpc$^{-1}$ to
$k=10^7$Mpc$^{-1}$, however we plotted the behavior from zero just
to see the behavior. As it can be seen from the left plot of Fig.
\ref{plot5} the circular polarization function $\Pi_{k\gg a H}(k)$
is non-zero from quite small $k$ while from , from
$k=0.08$Mpc$^{-1}$ it becomes constant and equal to $\Pi_{k\gg a
H}(k)=-1$, however we omitted the rest of the plot, because after
$k=0.08$Mpc$^{-1}$ the circular polarization function is constant.
Thus the subhorizon modes ar highly polarized, since $\Pi_{k\gg a
H}(k)\neq 0$. The same applies for the superhorizon modes, and in
the right plot of Fig. \ref{plot5} we present the behavior of the
circular polarization function $\Pi_{k\ll a H}(k)$ for modes that
were superhorizon during the first stages of inflation. As it can
be seen these modes are highly polarized too, but for these modes,
the wavenumber must not exceed $k\simeq 0.08$Mpc$^{-1}$, because
these modes are relevant only to CMB experiments.

\begin{figure}
\centering
\includegraphics[width=20pc]{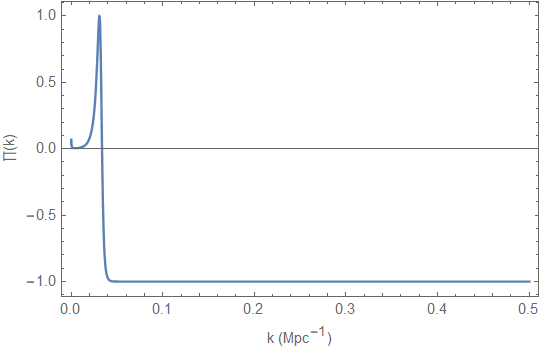}
\caption{The circular polarization function $\Pi_{k\ll a
H}(k)=\frac{|h_{L}^{k\ll a H}(k)|^2-|h_{R}^{k\ll a
H}(k)|^2}{|h_{L}^{k\ll a H}(k)|^2+|h_{R}^{k\ll a H}(k)|^2}$ for
the superhorizon modes, as function of the wavenumber of the modes
for the kinetic axion $R^2$ model.}\label{plot6}
\end{figure}

Now, let us turn our focus on the kinetic axion $R^2$ model. In
this case, analytic calculations are not possible for the
subhorizon case, thus even though we can obtain the superhorizon
solutions, we are not able to determine the integration constants.
However, just for the sake of completeness, we shall derive the
analytic solutions for superhorizon modes, and by using arbitrary
values for the constants of the integration, we shall demonstrate
that indeed the circular polarization function $\Pi(k)$ is
non-trivial in this case too. Recall that in the kinetic axion
model, $\dot{\phi}\sim a^{-3}$, and also $\ddot{\phi}\simeq
H\dot{\phi}$. Hence it is apparent that different terms of the
derivatives of the Chern-Simons coupling function $\nu (\phi)$
dominate the evolution eventually, as we now demonstrate. Let us
quote the evolution differential equations for the left and right
handed polarization modes at this point, and we show how these are
simplified eventually. With regard to the left-handed modes, the
evolution equation of the superhorizon modes is,
\begin{equation}\label{superhorizokineticleft}
\ddot{u}_L(t)+\dot{u}_L(t) \left(\frac{k \dot{a}(t) \dot{\nu}(t)-k
a(t) \ddot{\nu}(t)}{12 \alpha H_0^2 a(t)^2-k a(t) \dot{\nu}(t)}+3
H_0\right)=0\, ,
\end{equation}
while the right-handed modes satisfy,
\begin{equation}\label{superhorizokineticright}
\ddot{u}_R(t)+\dot{u}_R(t) \left(\frac{-k \dot{a}(t)
\dot{\nu}(t)+k a(t) \ddot{\nu}(t)}{12 \alpha H_0^2 a(t)^2-k a(t)
\dot{\nu}(t)}+3 H_0\right)=0\, .
\end{equation}
Now regarding the term in the denominator of the second term in
both the evolution equations, the term $12 \alpha H_0^2 a(t)^2$ is
dominant over $k a(t) \dot{\nu}(t)$ for the values of $k$
corresponding to superhorizon modes during inflation. Also, due to
the fact that for the kinetic axion we have $\ddot{\phi}\simeq
H\dot{\phi}$, the term $k a(t) \ddot{\nu}(t)$ in the numerator of
the second term in both the evolution equations is subdominant
compared to the term $k a(t) \ddot{\nu}(t)$, thus the evolution
equations for the left-handed modes becomes,
\begin{equation}\label{superhorizokineticleftsimp}
\ddot{u}_L(t)+\dot{u}_L(t) \left(\frac{-k a(t) \ddot{\nu}(t)}{12
\alpha H_0^2 a(t)^2}+3 H_0\right)=0\, ,
\end{equation}
while the volution equation for the right-handed modes is
simplified as follows,
\begin{equation}\label{superhorizokineticrightsimplified}
\ddot{u}_R(t)+\dot{u}_R(t) \left(\frac{\dot{\nu}(t)+k a(t)
\ddot{\nu}(t)}{12 \alpha H_0^2 a(t)^2}+3 H_0\right)=0\, .
\end{equation}
Regarding the conventions for the form of the Chern-Simons
coupling function $\nu (\phi)$, a complete study of the
Chern-Simons extended axion $R^2$ gravity is lacking for the
kinetic axion case, we intend to do this in a future work. Thus we
shall use for simplicity the conventions of the misaligned axion
case, for the sake of the argument. The qualitative picture is not
expected to dramatically change, when the conventions on the free
parameters are changed though, and this justifies our qualitative
approach here. Thus, using the conventions of the canonical
misalignment axion case, for a de Sitter background evolution, the
differential equations above can be solved analytically, with the
left handed solution being,
\begin{equation}\label{leftsolaxionsimplkinet}
u_L(t)=\mathcal{C}_1+\frac{12 \alpha  \mathcal{C}_2 H_0
e^{-\frac{\delta k e^{-3 H_0 t}}{36 \alpha H_0^2}}}{\delta k}\, ,
\end{equation}
while the right-handed solution is,
\begin{equation}\label{rightsolaxionsimplkinet}
u_R(t)=\mathcal{C}_3+\frac{12 \alpha  \mathcal{C}_4 H_0
e^{-\frac{\delta k e^{-3 H_0 t}}{36 \alpha H_0^2}}}{\delta k}\, ,
\end{equation}
where $\mathcal{C}_1$, $\mathcal{C}_2$, $\mathcal{C}_3$ and
$\mathcal{C}_4$ are arbitrary integration constants. As it can be
seen, both the solutions (\ref{leftsolaxionsimplkinet}) and
(\ref{rightsolaxionsimplkinet}) describe constant modes after
horizon crossing, which both contain an exponentially decaying
part. For this case, it is not possible to obtain the analytic
values for these constants, because we do not know the analytic
solutions for the subhorizon modes. Thus, just to see the
qualitative behavior of the circular polarization function
$\Pi(k)$, we shall take these to be of the order of unity. We
expect the overall qualitative behavior of the circular
polarization function will not be affected dramatically by the
actual values of the integration constants, however a formal
treatment of the problem requires the exact values of the
constants. Thus, using the same numerical conventions as in the
canonical misalignment axion case, in Fig. \ref{plot6} we plot the
circular polarization function $\Pi_{k\ll a
H}(k)=\frac{|u_{L}^{k\ll a H}(k)|^2-|u_{R}^{k\ll a
H}(k)|^2}{|u_{L}^{k\ll a H}(k)|^2+|u_{R}^{k\ll a H}(k)|^2}$ for
the superhorizon modes, in the $R^2$ kinetic misalignment axion
case.
As it can be seen from Fig. \ref{plot6}, the circular polarization
function is non-trivial in the kinetic axion $R^2$ model,
regarding the superhorizon modes. However, the behavior obtained
for the kinetic axion $R^2$ model is only a qualitative one,
because the correct treatment requires the analytic solutions for
the subhorizon modes, in order to correctly evaluate the arbitrary
integration constants. However, we do not expect that the overall
qualitative picture will dramatically change. The overall
conclusion for both the kinetic and canonical misalignment $R^2$
axion models is that the tensor modes, both subhorizon and
superhorizon modes are highly polarized. Thus if in future
gravitational wave experiments, two signals of stochastic
gravitational wave background are found, for the same frequency
range, this will be a smoking gun for the presence of Chern-Simons
terms in the inflationary Lagrangian. The axion $R^2$ models we
studied in this paper are quite appealing phenomenologically,
since, apart from the remnant chirality these generate, and the
viable inflationary era, they also provide a refined solution for
the dark matter problem. This is because the axion starts to
oscillate when $m_a\sim H$ and for all eras for which $m_a\gg H$,
and its energy density redshifts as $\rho_a\sim a^{-3}$, thus it
redshifts as dark matter. The difference between the two models is
the time at which the oscillations begin, but beyond that
difference, during the post-reheating era, the two models are
basically the same. Before the reheating, the kinetic axion $R^2$
model might be more interesting because this model might lead to a
lower reheating temperature. The physics of this model shall be
studied elsewhere.

\section{Conclusions}

In this paper we studied the chirality of primordial gravitational
waves in the context of Chern-Simons axion $f(R)$ gravity.
Specifically, the $f(R)$ gravity was chosen to be the $R^2$ model,
and we considered two mainstream axion field theory models, the
canonical misalignment axion model and the kinetic axion model.
The presence of the Chern-Simons term in the context of
Einstein-Hilbert gravity ensures the chirality of the primordial
gravitational waves, and this was the focus of this work, to check
whether this remains true in the case of Chern-Simons $f(R)$
gravity, with the scalar field being the axion. As we showed,
since the axion and the corresponding Chern-Simons term  do not
significantly affect the background evolution, the $R^2$ model
completely determines the background evolution. However, the
Chern-Simons term affects the tensor perturbations explicitly, and
it modifies the evolution of the two distinct polarization modes.
Since we were interested in quantifying the chirality
modifications caused by the Chern-Simons term we aimed to solve
analytically the evolution equations of each polarization modes
for each of the axion models. We considered subhorizon modes and
superhorizon modes, each of which are probed or will be probed
distinctly by the future gravitational wave experiments and the
current and future CMB-based experiments. For the case of the
misalignment axion we were able to find analytic expressions for
both the superhorizon and subhorizon modes, and we were able to
find all the integration constants. Accordingly, we calculated the
circular polarization function for each of the subhorizon and
superhorizon modes as a function of the wavenumber, and as we
showed the modes are strongly chiral. In the case of the kinetic
axion, the analytical study of the subhorizon modes was
impossible, thus for the sake of the argument, we used the
conventions of the misalignment axion case and we also showed that
the spectrum is also chiral. However, the essential features of
the kinetic axion $f(R)$ gravity and its Chern-Simons extension
are needed, which are lacking from the literature, and we aim to
address in a future work.

\section*{Acknowledgments}

This work was supported by MINECO (Spain), project
PID2019-104397GB-I00 (S.D.O). This work by S.D.O was also
partially supported by the program Unidad de Excelencia Maria de
Maeztu CEX2020-001058-M, Spain.

\end{document}